\documentclass{Interspeech}
\DeclareUnicodeCharacter{202F}{\,}
\usepackage{graphicx} 
\usepackage{tabularx} 
\usepackage{booktabs} 
\usepackage{pifont} 
\usepackage{float} 
\usepackage{stfloats} 
\usepackage{makecell} 
\usepackage{multirow}
\usepackage{amsmath}
\usepackage{threeparttable}
\usepackage{booktabs} 
\usepackage{makecell} 
\usepackage{bm}
\usepackage{mathrsfs}
\usepackage[mathscr]{euscript}
\usepackage{cite}
\usepackage{enumitem}
\usepackage{colortbl}
\usepackage[table]{xcolor}

\interspeechcameraready

\title{Enhancing Target Speaker Extraction with Explicit Speaker Consistency Modeling   }

\author[affiliation={1}]{Shu}{Wu}
\author[affiliation={1}]{Anbin}{Qi}
\author[affiliation={1}]{Yanzhang}{Xie}

\author[affiliation={1,2}]{Xiang}{Xie$^{*}$}

\affiliation{School of Information and Electronics}{Beijing institute of Technology }{China}
\affiliation{}{Beijing institute of Technology, Zhuhai}{China}
\email{wushu@bit.edu.cn, 3220220692@bit.edu.cn, x1eyzh@163.com,xiexiang@bit.edu.cn}

\usepackage{comment}

\begin{document}

\maketitle
\renewcommand{\thefootnote}{\fnsymbol{footnote}}
\footnotetext[1]{Corresponding author.}
\renewcommand{\thefootnote}{\arabic{footnote}}
\begin{abstract}

Target Speaker Extraction (TSE) uses a reference cue to extract the target speech from a mixture. In TSE systems relying on audio cues, the speaker embedding from the enrolled speech is crucial to performance. However, these embeddings may suffer from speaker identity confusion. Unlike previous studies that focus on improving speaker embedding extraction, we improve TSE performance from the perspective of speaker consistency. In this paper, we propose a speaker consistency-aware target speaker extraction method that incorporates a centroid-based speaker consistency loss. This approach enhances TSE performance by ensuring speaker consistency between the enrolled and extracted speech.  In addition, we integrate conditional loss suppression into the training process.  The experimental results validate the effectiveness of our proposed methods in advancing the TSE performance. A speech demo is available online.\footnote{https://sc-tse.netlify.app/}

\end{abstract}

\section{Introduction}

In real-world scenarios, we navigate complex acoustic environments filled with overlapping speech and diverse background noise, such as music or machine-generated sounds. The challenge of isolating the target speech while ignoring other interferences is known as the \textit{cocktail party problem} \cite{cherry1954some}. Humans achieve this through selective attention \cite{bronkhorst2015cocktail}, focusing on a specific speaker while filtering out distractions. Target Speaker Extraction (TSE) tackles this challenge by leveraging audio \cite{speakerbeam,jansky2020adaptive}, visual \cite{Ephrat_Mosseri_Lang_Dekel_Wilson_Hassidim_Freeman_Rubinstein_2018, hershey2001audio}, or spatial \cite{Gu_Chen_Zhang_Zheng_Xu_Yu_Su_Zou_Yu_2019} cues as references.

Emerging around 2017, TSE initially relied on pre-trained speaker encoders from verification tasks to extract speaker embeddings. SpeakerBeam~\cite{speakerbeam} introduced single-channel masking and multichannel beamforming, while \cite{gabrys2022voice} focused on ASR integration and low-latency streaming. Extensions like speaker inventory~\cite{Xiao_Chen_Yoshioka_Erdogan_Liu_Dimitriadis_Droppo_Gong_2019} adapted TSE for multi-speaker scenarios. However, models trained on limited datasets often struggle with speaker confusion—mistakenly extracting the wrong speaker or generating fragmented utterances~\cite{Target-Conf}.

Different from using a pre-trained speaker encoder, the current popular TSE models are typically trained jointly with an auxiliary module that generates speaker embeddings for the target speaker \cite{Ge_Xu_Wang_Chng_Dang_Li_2020,liu2023x}.  However, these jointly trained encoders often struggle with speaker discrimination in TSE tasks \cite{zhang2023speaker}. Zhao et al. \cite{Target-Conf} link this issue to high acoustic similarity and the encoder's limited ability to capture speaker-specific traits.

To enhance robustness and generalization, recent research has explored diversifying enrollment data \cite{on} and integrating self-supervised models \cite{lin2024selective,peng2024target} to capture richer and more diverse speaker representation. Some studies focus on frame-level embeddings \cite{contra,real}, which aim to improve speaker discrimination by utilizing fine-grained features. Meanwhile, other methods bypass speaker encoders altogether, directly fusing speaker information at the spectrogram level \cite{hu2024smma,yang2024target}.

Previous studies mainly aimed at enhancing speaker embedding quality to reduce speaker confusion and improve TSE performance. However, these approaches overlook a fundamental aspect of TSE: while the reference and estimated target speech differ in content, they originate from the same speaker. Moreover, maintaining stable speaker characteristics in the extracted speech is crucial for downstream tasks. To our knowledge, no prior work has systematically explored the impact of speaker representation consistency between the reference and extracted speech on TSE performance.

In this paper, we thoroughly investigate the impact of speaker representation consistency between reference and extracted speech on the performance of the TSE system.  The main contributions of this work include: 1)We propose a centroid-based speaker consistency loss for TSE system. This approach not only significantly enhances the consistency between extracted and target speech but also improves the overall speech quality.
2) We integrate conditional loss suppression into train progress, enhancing the overall performance of the TSE system. 
3) We conduct a comprehensive evaluation to validate the effectiveness of the proposed methods for target speaker extraction. The proposed methods enhance overall performance, achieving state-of-the-art results. The ablation study further confirmed the effectiveness of the methods. Meanwhile, experiments on different backbones and out-of-domain datasets demonstrate our methods' strong generalization and robustness.

This paper is organized as follows. In Section 2, we discuss the details of our proposed methods. In Section 3, we report the experiment setup. In Section 4, we summarize the results. Finally, Section 5 concludes the discussion and future work.

\section{Proposed methods}

\subsection{Model architecture overview}
\begin{figure*}[htbp]
  \centering
  \includegraphics[width=\linewidth]{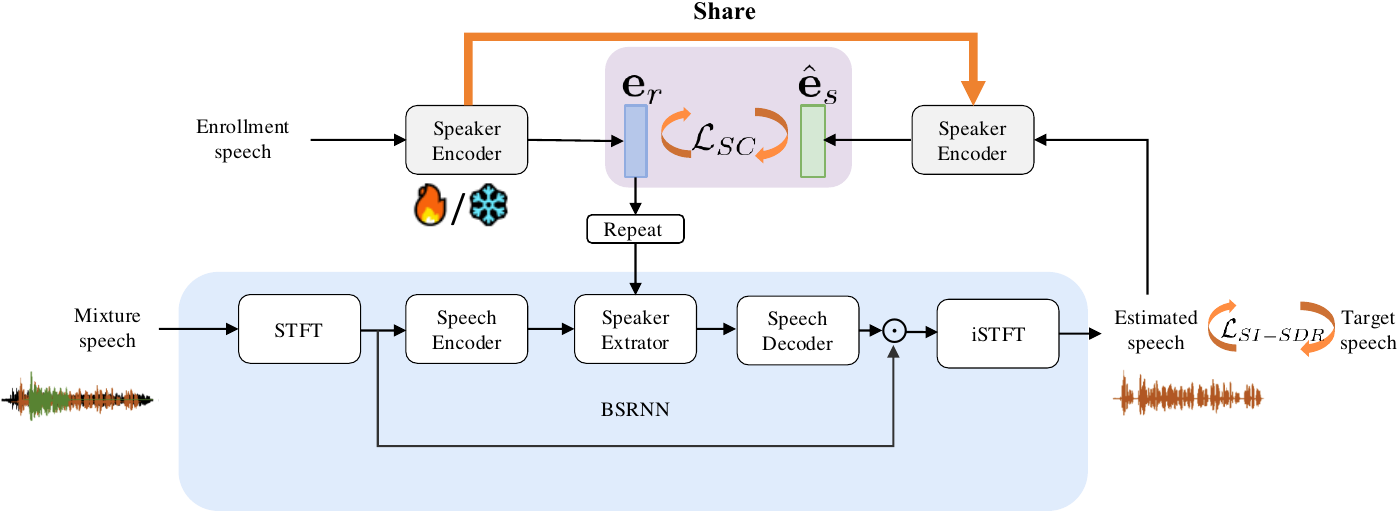}
   \vspace{-7mm} 
  \caption{The architecture of the proposed speaker consistency-aware target speaker extraction model}
  \label{fig:secs model}
  \vspace{1em} 
  
  \vspace{-5mm} 
\end{figure*}

We adopt a standard speaker extraction pipeline (Figure \ref{fig:secs model}), comprising a speaker encoder, a separation module, and the proposed speaker consistency loss module. The Band-Split RNN (BSRNN) \cite{bsrnn} serves as the separator backbone. The speaker encoder is either pre-trained or jointly trained with the separator. The speaker consistency loss module is highlighted. The speaker encoder extracts $\mathbf{e}_{r}$ from enrollment speech and $\hat{\mathbf{e}}_{s}$ from extracted speech, sharing  the same parameters.

\subsection{Centroid-based speaker consistency loss}

\subsubsection{Speaker consistency in TSE}
Speaker similarity is commonly used to assess the degree of similarity between the speech of different speakers. In the field of Text-To-Speech (TTS) study\cite{tts}, speaker encoder cosine similarity (SECS) is utilized to evaluate the speaker consistency between the synthesized speech and the target speaker's speech. In the TSE task, we similarly aim to extract speech that retains a high degree of speaker similarity to the target speech from the same speaker.  The SECS (with values ranging from -1 to 1) for the TSE task can be defined as follows:
\begin{equation}
\begin{aligned}
\mathit{SECS} &= \mathit{cos}\left(\mathbf{e}_{r}, \hat{\mathbf{e}}_{s}\right) \\
&= \mathit{cos}\left(E_\theta(r), E_\theta(\hat{s})\right)
\end{aligned}
\label{eq:SECS}
\end{equation}
where \(E_\theta(\cdot)\) denotes the speaker encoder, either pretrained or jointly trained with the speech extraction module. $\hat{\mathbf{e}}_{r}$ is the speaker embedding  from the enrolled speech $r$ , $\mathbf{e}_{s}$ is the speaker embedding  from the extracted speech $\hat{s}$ , and the $\mathit{cos}(\cdot)$ denotes the cosine similarity function.

\subsubsection{Speaker consistency loss}
As defined in the TSE  task, the input reference speech and the estimated target speech, while differing in content, are both uttered by the same speaker. Given this premise, it is reasonable to expect that when these two signals are processed through the speaker encoder, the resulting speaker embeddings should become progressively more similar. This inherent intuition forms the basis for the formulation of the speaker  consistency loss ($\mathcal{L}_{SC}$), which can be mathematically expressed as follows:
\begin{equation}
    \mathcal{L}_{SC} = 1 - SECS
\end{equation}

\subsubsection{Centroid-based speaker consistency loss}
\label{centr}
Prototype learning has emerged as a promising approach in speaker recognition\cite{prototype}, leveraging class prototypes to represent the central characteristics of speaker embeddings. Wan et al.\cite{speakercentroid} refer to these class prototypes as the speaker centroids, and the utilization of the speaker centroid significantly enhances the system's robustness. Inspired by this, we propose Centroid-based speaker consistency loss.

 To obtain the centroids, we first utilize a pretrained speaker encoder, consistent with the architecture in Figure \ref{fig:secs model},  to extract the utterance-level speaker embedding $\mathbf{e}_{i,k}^{U}$ for all utterances from  speaker $\mathbf{i}$:

\begin{align}
\mathbf{e}_{i,k}^{U} &= E_\theta\left(\mathbf{x}_{i,k}\right)
\end{align}
where k is the utterance index. The speaker centroid for the i-th speaker $\mathbf{e}_{i}^{C}$  are derived by averaging all these utterance-level embeddings:
\begin{align}
\mathbf{e}_{i}^{C} &= \frac{1}{K}\sum_{k=1}^{K}\mathbf{e}_{i,k}^{U},
\end{align}

After obtaining $\mathbf{e}_{i}^{C}$, the definition of Centroid-based speaker consistency loss is as follows:

\begin{align}
\mathcal{L}_{C-SC} = -\log\frac{\exp(\mathit{cos}(\hat{\mathbf{e}}_{s}, \mathbf{e}_{i^{T}}^{C}))}{\sum_{i=1}^{N} \exp(\mathit{cos}(\hat{\mathbf{e}}_{s}, \mathbf{e}_{i}^{C}))}
\label{lsecs}
\end{align}
where  $ \mathbf{i^{T}}$ is the target speaker index. N is the speaker number. When using $\ensuremath{\mathcal{L}_{C-SC}}$, the current reference speech speaker embedding does not participate in the loss computation but still serves as the reference cue of the speech separator module.

\subsection{Training}\label{sec:figures}
\subsubsection{Loss function}\label{loss}
SI-SDR \cite{sdr} loss is used to measure the quality between the estimated and clean target speech. When the speaker encoder is learnable, we add a cross-entropy (CE) loss for speaker classification. Then we integrate the proposed centroid-based speaker consistency loss to obtain the final loss:
\begin{equation}
    \mathcal{L} = \left(1-\beta -\lambda \right)\mathcal{L}_{SI-SDR}+\beta  \mathcal{L}_{CE}+\lambda \mathcal{L}_{C-SC}
    \label{eq:final loss}
\end{equation}
where $\beta$ and $\lambda$ is the scaling factor, when the speaker encoder is frozen, we set $\beta$ to 0; otherwise, we set $\beta$ to 0.1. When  \ensuremath{\mathcal{L}_{C-SC}} is not adopted, we set $\lambda$ to 0; otherwise, we set $\lambda$ to 0.1.

\subsubsection{Conditional loss suppression}\label{CLS}
In our TSE architecture, the speaker encoder, separator model, and speaker consistency module serve distinct roles: the speaker encoder extracts target speaker embeddings, the separator extracts high-quality speech, and the speaker consistency module ensures speaker representation consistency. A high SECS value indicates strong speaker similarity between the enrolled and extracted speech, but excessive SECS optimization may lead to overfitting and degrading separator performance. To mitigate this, we propose SECS Conditional Loss Suppression (CLS), which deactivates speaker consistency loss once it surpasses a predefined threshold.  We formalize this strategy by first defining the CLS function $f_{C}(\cdot)$ as follows: 
\begin{equation}
f_{C}\left( x \right) =
\begin{cases}
x, & \text{if } SECS \leq \omega\, \\
0, & \text{otherwise.}
\end{cases}
\end{equation}

Where $\omega$ is the SECS threshold, which can remain fixed or vary during training. In this work, $\omega$ decreases linearly from 1.0 to 0.8. With the CLS method, the loss function \eqref{eq:final loss} is reformulated as follows:
\begin{equation}
    \mathcal{L}_{C} = \left(1-\beta -\lambda \right)\mathcal{L}_{SI-SDR}+\beta \mathcal{L}_{CE}+ f_{C}\left( \lambda\mathcal{L}_{C-SC} \right)
\end{equation}

The hyperparameters of the formula are consistent with those in \eqref{eq:final loss}.

\section{Experiments}
\begin{table*}[h!]
  \caption{The results of BSRNN with proposed methods on Libri2Mix}
  
  \vspace{-6mm} 
    \begin{flushleft}
    \small
    Performance of BSRNN with our proposed methods. The experiments are grouped by two speaker encoder structures, ECAPA-TDNN and ResNet34, and two training strategies, pretrained and joint. The upward arrow $\uparrow$ indicates better performance with higher values.
  \end{flushleft}
  \vspace{-3mm} 
  \label{tab:The results of BSRNN with SECSL}
  \centering
  \begin{tabular*}{\textwidth}{@{\extracolsep{\fill}}c c| c| c | c | c | c | c | c | c | c  c}
    \toprule
    \multicolumn{2}{c|}{\makecell{\textbf{Speaker}\\\textbf{Model}}} & 
    
    \multicolumn{1}{c|}{\makecell{\textbf{Train}\\\textbf{Method}}} & 
    \multicolumn{1}{c|}{\ensuremath{\mathcal{L}_{C-SC}}} & 
    \multicolumn{1}{c|}{\textbf{CLS}} & 
    \multicolumn{1}{c|}{\makecell{\textbf{SI\_SDR}\\\textbf{/dB$\uparrow$}}} & 
    \multicolumn{1}{c|}{\textbf{Acc. / \%}$\uparrow$} & 
    \multicolumn{1}{c|}{\textbf{Sim. / \%}$\uparrow$} & 
    \multicolumn{1}{c|}{\textbf{SDR / dB}$\uparrow$} & 
    \multicolumn{1}{c|}{\textbf{PESQ}$\uparrow$} & 
    \multicolumn{1}{c}{\textbf{STOI}$\uparrow$} \\
    \midrule 
    
    \multicolumn{2}{c|}{\multirow{6}{*}{\textbf{ECAPA}}} & \multirow{3}{*}{\quad\textbf{Pretrained\quad 
        }}  &   & &  $13.34$ & $91.08$  &$84.28$ & $14.80$ & $2.72$ & $90.34$  \\
    \multicolumn{2}{c|}{} & & \checkmark & & $13.85$ & $92.10$ & \pmb{86.92} & $14.85$ & $2.74$ & \pmb{92.95} \\
    \multicolumn{2}{c|}{} & & \checkmark & \checkmark & \pmb{14.29} & \pmb{95.15} & $86.83$ & \pmb{15.19} & \pmb{2.77} & $91.06$ \\
    \cline{3-11}
    \multicolumn{2}{c|}{} & \multirow{3}{*}{\textbf{Joint}}     &        & & $13.98$ & $93.85$  &$86.62$ & $15.16$ & $2.67$ & $91.95$  \\
    \multicolumn{2}{c|}{}  & &       \checkmark   &     & $14.58$ & $95.77$  &\pmb{87.98} & $15.36$ & $2.73$ & \pmb{92.18}  \\
    \multicolumn{2}{c|}{} & & \checkmark & \checkmark & \pmb{14.63} & \pmb{96.70} & $87.69$ & \pmb{15.38} & \pmb{2.75} & \pmb{92.18} \\
    \midrule
    \multicolumn{2}{c|}{\multirow{6}{*}{\textbf{ResNet34}}} & \multirow{3}{*}
    {\quad\textbf{Pretrained\quad 
        }}  & & & $13.21$ & $92.26$ & $85.84$ & $14.58$ & $2.71$ & $90.08$ \\
    \multicolumn{2}{c|}{} & & \checkmark & & $13.64$ & $93.82$ & \pmb{86.81} & $14.68$ & \pmb{2.73} & $90.33$ \\
    \multicolumn{2}{c|}{} & & \checkmark & \checkmark & \pmb{14.10} & \pmb{94.36} & $86.57$ & \pmb{14.72} & $2.71$ & \pmb{90.82} \\
    \cline{3-11}
    \multicolumn{2}{c|}{} & \multirow{3}{*}{\textbf{Joint}}     &        & & $14.06$ & $95.07$  &$86.06$ & $14.87$ & $2.70$ & $91.49$  \\
    \multicolumn{2}{c|}{}  & &       \checkmark   &     & $14.54$ & $95.85$  &\pmb{87.48} & $15.37$ & $2.71$ & $92.10$  \\
    \multicolumn{2}{c|}{} & & \checkmark & \checkmark & \pmb{14.75} & \pmb{96.17} & $86.62$ & \pmb{15.40} & \pmb{2.76} & \pmb{92.17} \\
    
    \bottomrule
  \end{tabular*}
  \vspace{-3mm}

\end{table*}
\subsection{Dataset}

 We use the clean Libri2Mix \cite{libri2mix} dataset with two-speaker mixtures. TSE models are trained on the 100-hour subset (13900 utterances, 251 speakers), with validation and test sets containing 3,000 utterances from 40 non-overlapping speakers. Each sample is used twice for different speaker extractions by alternating enrollment. The dataset features fully overlapped audio (minimum version) at 16 kHz.

\subsection{Training details}

In the TSE model, BSRNN serves as the speech separator. The speech encoder splits the frequency bands as follows: below 1.5 kHz (100 Hz bandwidth), 1.5–3.5 kHz (200 Hz bandwidth), 3.5–6 kHz (500 Hz bandwidth), and 6–8 kHz as a single subband, resulting in 32 subbands with a feature dimension of 128. The speaker separator uses a 192-dimensional bidirectional LSTM, repeated 6 times in the BSRNN. ResNet34\cite{resnet} and ECAPA-TDNN\cite{ecapa} from the WeSpeaker toolkit\footnote{https://github.com/wenet-e2e/wespeaker}\cite{wespeaker}, pretrained on 5,994 speakers from VoxCeleb2\cite{voxceleb}, serve as speaker encoders in pretrain mode and for obtaining $\mathbf{e}_{i,k}^{U}$ in Section \ref{centr}. In joint training mode, these models are trained from scratch with the speech separator model. Our training is based on the WeSep \cite{wang2024wesepscalableflexibletoolkit} framework, with reference to some of its configurations. TSE models are trained for 150 epochs on 3-second segments using the Adam optimizer, starting at a learning rate of 1e-3, and decaying to 2.5e-5. The last 5 checkpoints are averaged for inference.

\subsection{Evaluation metrics}

 We use common objective metrics, including SI-SDR\cite{sdr}, SDR \cite{sdr}, PESQ \cite{pesq}, and STOI\cite{stoi}. Besides, we consider the samples with SI-SDRi larger than 1 dB as the successful extraction. The percentage of these samples is measured for the accuracy (Acc.) of extraction. Notably, Speaker Similarity (Sim.) evaluates the speaker consistency between the extracted and target speech, computed using ZS-TTS-Evaluation toolkit,\footnote{https://github.com/Edresson/ZS-TTS-Evaluation}\cite{casanova24_interspeech}with ECAPA2\cite{ecapa2}serving as the speaker encoder.

\section{Results}

\subsection{Comparative studies with proposed methods}
Table \ref{tab:The results of BSRNN with SECSL} presents the performance of BSRNN with proposed methods on  Libri2Mix. Four comparative studies were conducted using ECAPA-TDNN and ResNet34 models under pretrained or joint training modes, without utilizing the speaker centroid. In each case, the BSRNN model combined with a speaker encoder served as the baseline.  From this table, we can draw the following conclusions:
\begin{enumerate}[label=\arabic*., leftmargin=*, itemsep=0.1em, labelsep=0.5em]
    \item   The centroid-based speaker consistency loss improve both Sim. and other metrics in all scenarios. With pretrained ECAPA-TDNN, SI-SDR improved by 0.51 dB, accuracy by 1.02\%, and Sim. by 2.64\%, alongside consistent gains across metrics. Joint training show similar trends, with SI-SDR improving by 0.60 dB, accuracy by 1.92\%, and Sim. by 1.36\%. 
    \item The centroid-based speaker consistency loss demonstrates strong generalization across speaker encoders. For ResNet34 in pretrained mode,   SI-SDR improved by 0.43 dB, accuracy by 1.56\%, and Sim. by 0.97\%. Similar gains are observed in joint training, with comparable increases in all metrics.
    \item  The CLS strategy  enhances TSE system performance, improving SI-SDR, accuracy, and other metrics in all scenarios. However, a slight decrease in the Sim. metric was observed, though CLS still outperformed the baseline. This may result from the suppressive effect of CLS on centroid-based speaker consistency loss.
    \item  Overall, by combining centroid-based speaker consistency loss with the CLS strategy, our proposed methods lead to improvements in the TSE system's performance.
\end{enumerate}

\begin{table}[th]
  \caption{Ablation results on  Libri2Mix}
  \vspace{-6mm} 
  \begin{flushleft}
    \small
    SC-BSRNN represents the speaker consistency-aware BSRNN with our proposed\ensuremath{\mathcal{L}_{C-SC}} and CLS. The row highlighted in light gray presents the results using the  \ensuremath{\mathcal{L}_{SC}} without CLS. All models use pretrained ECAPA-TDNN as a speaker model.
  \end{flushleft}
  \vspace{-3mm} 
  
  \label{tab:speaker centroid}
  \centering
  \begin{tabular}{@{}c@{\extracolsep{\fill}}c c c c @{}}
    \hline
    \multicolumn{2}{c}{\textbf{Model}} & 
    \makecell{\textbf{SI\_SDR}\\\textbf{/dB}} & 
    \makecell{\textbf{Acc.}\\\textbf{/ \%}} & 
    \makecell{\textbf{Sim.}\\\textbf{/ \%}} \\
    \hline
    \textbf{SC-BSRNN}  &  & \pmb{14.29} & \pmb{95.15} & \pmb{86.83} \\
    w/o speaker centroid &  & $14.24$ & $94.93$ & $86.66$ \\  
    
    \cellcolor{gray!25} w/o speaker centroid , CLS
 &  & \cellcolor{gray!25}$13.77$ & \cellcolor{gray!25}$91.42$ & \cellcolor{gray!25}$86.75$ \\
    \hline
    BSRNN (baseline) &  & $13.34$ & $91.08$ & $84.28$ \\
    \hline
  \end{tabular}

  \vspace{-2mm} 
\end{table}

\subsection{Ablation study}  
Table \ref{tab:speaker centroid} illustrates the impact of integrating the speaker centroid and CLS into the speaker consistency loss on TSE performance. The baseline system employed BSRNN with a pretrained ECAPA-TDNN. Results show that the speaker centroid consistently improves SI-SDR, Accuracy, and speaker similarity by mitigating over-reliance on the enrollment speech embedding, ensuring a more balanced speaker representation. This enhances the system’s robustness and generalization across speech segments from the same speaker. Additionally, CLS improves SI-SDR and Accuracy but introduces slight trade-offs in speaker similarity.  

Notably, the model trained with the original speaker consistency loss \ensuremath{\mathcal{L}_{SC}} surpasses the baseline across all metrics, demonstrating its effectiveness in enhancing TSE performance, regardless of the inclusion of the speaker centroid and CLS.

\begin{table}[th]  

  \caption{Comparison with other methods on Libri2Mix. Results for other methods are cited from original papers and  \cite{zhang2024multilevelspeakerrepresentationtarget}}  
  \vspace{-2mm}
  \label{tab:sot}  
  \centering  
  \small 
  \begin{tabular}{@{}c@{\extracolsep{1pt}}c c c c@{}}  
    \toprule  
    \textbf{Model}  & \makecell{\textbf{Speaker}\\\textbf{Model}}  &  \makecell{\textbf{Training}\\\textbf{Method}}& \makecell{\textbf{SI\_SDRi}\\\textbf{/dB}}      & \makecell{\textbf{Acc.}\\\textbf{/\%}} \\
    \midrule  
    SpeakerBeam\cite{speakerbeam} & ResNet & Joint & $13.03$ & $95.20$ \\
    SpEx+\cite{Ge_Xu_Wang_Chng_Dang_Li_2020} & ResNet & Joint &$13.41$  & - \\
    DPCCN \cite{dpcnn}& ConvNet &  Joint&$11.65$  &-  \\
    MC-SpEx \cite{MC-SpEx}&  ResNet & Joint &$14.61$  &  -\\
    Target-Conf\cite{Target-Conf} & ResNet  &Joint  &$13.88$  &- \\
    X-T-TasNet \cite{speakercentroid}& d-vector & Pretrained & $13.48$ &  $95.3$\\
    \midrule  
    \makecell{SSL-TD-\\SpeakerBeam \cite{ssl}} & \makecell{ResNet+\\WavLM} & Pretrained & $14.65$ & $96.1$ \\
    \midrule  
    \multirow{4}{*}{\textbf{SC-BSRNN}} & \multirow{2}{*}{ECAPA} & Pretrained & $14.29$ & $95.15$ \\
    & & Joint & $14.63$ & \pmb{96.70} \\
    \cline{2-5}  
    & \multirow{2}{*}{ResNet} & Pretrained & $14.10$ & $94.36$ \\
    & & Joint & \pmb{14.75} & $96.17$ \\
    \bottomrule  
  \end{tabular}

  \vspace{-5mm} 
\end{table}

\subsection{Comparison with other methods}
Table \ref{tab:sot} presents a comparison of our results with other methods  on the Libri2Mix dataset. Our proposed model achieves the best performance among all  TSE systems in terms of SI-SDRi and accuracy. We specifically include a comparison with recent studies that focus on extracting speaker features using large models trained through Self-Supervised Learning (SSL). These approaches not only leverage pre-trained speaker encoders but also utilize SSL-based speech models for speaker information extraction. In contrast, our proposed method achieves performance comparable to these approaches while maintaining the standard TSE pipeline, offering a distinct advantage in terms of overall model size.

\subsection{Generalization of proposed methods }
Table \ref{generation1} demonstrates the generalization ability of our proposed methods across different speech separator backbones. We conduct experiments on  Libri2Mix using two backbone models, DPCCN and TF-GridNet\cite{tf}. The results indicate that our proposed methods exhibit strong generalization to various backbones, significantly improving TSE performance on both architectures, which highlights the flexibility and adaptability of our approach in different settings.

Table \ref{generation2} demonstrates the generalization ability of our proposed methods on out-of-domain datasets. We train the models on the Libri2Mix-train-100 and VoxCeleb1\cite{vox} datasets and evaluate them on Libri2Mix and Aishell2Mix\footnote{https://github.com/jyhan03/icassp22-dataset}\cite{dpcnn}. The results show that SC-BSRNN consistently outperforms  both in the in-domain and the out-of-domain datasets, indicating the strong generalization capability of our proposed methods in unseen scenarios.

\begin{table}[th]
  
  \caption{Generalization  on different TSE backbones}
  \vspace{-6mm} 
  \begin{flushleft}
    \small
     Models with the "SC" prefix indicate the use of our proposed methods. TF-GridNet use pretrained ECAPA-TDNN model, DPCCN use the joint training method.
  \end{flushleft}
  \vspace{-3mm} 
  
  \label{generation1}
  \centering
  \begin{tabular}{@{}c@{\extracolsep{\fill}}c c c c @{}}
    \hline
    \multicolumn{2}{c}{\textbf{Model}} & 
    \makecell{\textbf{SI\_SDR}\\\textbf{/dB}} & 
    \makecell{\textbf{Acc.}\\\textbf{/ \%}} & 
    \makecell{\textbf{Sim.}\\\textbf{/ \%}} \\
    \hline
    \textbf{SC-DPCCN}  &  & \pmb{12.52} & \pmb{90.03} & \pmb{83.81} \\
    DPCCN &  & $11.65$ & $89.57$ & $83.06$ \\  
    \hline
    \textbf{SC-TF-GridNet}  &  & \pmb{12.97} & \pmb{90.21} & \pmb{84.65} \\
    
    TF-GridNet &  & $12.15$ & $89.77$ & $83.20$ \\
    \hline
  \end{tabular}

  \vspace{-2mm} 
\end{table}

\begin{table}[th]
  \vspace{-2mm}
  \caption{Generalization on out-of-domain dataset}
  \vspace{-3mm} 
  \label{generation2}
  \centering
  \begin{tabular}{@{}c@{\extracolsep{1pt}}c c c c@{}}
    \hline
    \multirow{2}{*}{\textbf{Model}} & \multirow{2}{*}{\textbf{Training Dataset}} & \multicolumn{2}{c}{\textbf{Evaluation (SI-SNR)}} \\
    \cline{3-4}
    & & Libri2Mix & Aishell2Mix \\
    \hline
    \textbf{SC-BSRNN} &\multirow{2}{*}{\makecell{Libri2Mix-\\train-100}}  & \pmb{14.29} & \pmb{5.89}  \\
    BSRNN & & $13.34$ & $5.51$  \\  
    \hline
    \textbf{SC-BSRNN} & \multirow{2}{*}{VoxCeleb1} & \pmb{16.47} & \pmb{10.36}  \\
    BSRNN & & $16.13$ & $10.12$  \\
    \hline
  \end{tabular}
  \vspace{-2mm} 
\end{table}

\section{Conclusion and future work}
In this paper, we propose a speaker consistency-aware target speaker extraction that integrates centroid-based speaker consistency loss and conditional loss suppression. Experiments demonstrate the effectiveness of our approach, showing substantial improvements not only in the speaker consistency between the extracted and target speech but also in the overall quality of the extracted speech.  Furthermore, the proposed methods exhibit strong generalization and robustness across different speech separator backbones and out-of-domain datasets. In future work, we plan to further investigate the impact of speaker consistency on TSE systems, apply it to more TSE models, and explore different integration strategies.

\bibliographystyle{IEEEtran}

\bibliography{mybib}

\end{document}